\documentclass[epj]{svjour}
\usepackage{graphics}
\begin{document}
\title{Trigonometric quark confinement potential of QCD traits.}
\author{C.\ B.\ Compean, M. Kirchbach}
%
\institute{Instituto de Fisica, UASLP,
Av. Manuel Nava 6, Zona Universitaria,\\
San Luis Potosi, SLP 78290, M\'exico
}
\date{Received: date / Revised version: date}
%
\abstract{
We make the case that the Coulomb-- plus  linear quark confinement potential
predicted by lattice QCD
is an approximation to the  exactly solvable
trigonometric Rosen-Morse potential
that has the property to interpolate between the Coulomb-- and
the infinite wells. We test the predictive power of this potential
in the description  of the nucleon  (considered 
as a quark-diquark system) and provide analytic expressions for
its mass spectrum and the proton electric 
form factor. We compare the results obtained in this fashion to 
data and find quite good  agreement. We obtain 
an effective  gluon propagator in closed form
as the Fourier transform of the
potential under investigation.  
} 
%
%

\maketitle



\def\s{\mbox{\boldmath$\displaystyle\mathbf{\sigma}$}}
\def\J{\mbox{\boldmath$\displaystyle\mathbf{J}$}}
\def\K{\mbox{\boldmath$\displaystyle\mathbf{K}$}}
\def\P{\mbox{\boldmath$\displaystyle\mathbf{P}$}}
\def\p{\mbox{\boldmath$\displaystyle\mathbf{p}$}}
\def\hp{\mbox{\boldmath$\displaystyle\mathbf{\widehat{\p}}$}}
\def\x{\mbox{\boldmath$\displaystyle\mathbf{x}$}}
\def\0{\mbox{\boldmath$\displaystyle\mathbf{0}$}}
\def\bv{\mbox{\boldmath$\displaystyle\mathbf{\varphi}$}}
\def\hbv{\mbox{\boldmath$\displaystyle\mathbf{\widehat\varphi}$}}

\def\bg{\mbox{\boldmath$\displaystyle\mathbf{\gamma }$}}

\def\bl{\mbox{\boldmath$\displaystyle\mathbf{\lambda}$}}
\def\br{\mbox{\boldmath$\displaystyle\mathbf{\rho}$}}
\def\1{\mbox{\boldmath$\displaystyle\mathbf{1}$}}
\def\bfhh{\mbox{\boldmath$\displaystyle\mathbf{(1/2,0)\oplus(0,1/2)}\,\,$}}

\def\mn{\mbox{\boldmath$\displaystyle\mathbf{\nu}$}}
\def\amn{\mbox{\boldmath$\displaystyle\mathbf{\overline{\nu}}$}}

\def\mne{\mbox{\boldmath$\displaystyle\mathbf{\nu_e}$}}
\def\amne{\mbox{\boldmath$\displaystyle\mathbf{\overline{\nu}_e}$}}
\def\rlh{\mbox{\boldmath$\displaystyle\mathbf{\rightleftharpoons}$}}

\def\wm{\mbox{\boldmath$\displaystyle\mathbf{W^-}$}}
\def\hh{\mbox{\boldmath$\displaystyle\mathbf{(1/2,1/2)}$}}
\def\h00h{\mbox{\boldmath$\displaystyle\mathbf{(1/2,0)\oplus(0,1/2)}$}}
\def\znbb{\mbox{\boldmath$\displaystyle\mathbf{0\nu \beta\beta}$}}



\newcommand{\csch}{\textrm{ csch }}
\newcommand{\sech}{\textrm{ sech }}
\newcommand{\arccot}{\textrm{ arccot }}
\newcommand{\arccoth}{\textrm{ arccoth }}
\newcommand{\e}{\textrm { e}}

\newcommand{\be}{\begin{eqnarray}}
\newcommand{\ee}{\end{eqnarray}}
\newcommand{\nn}{\nonumber}

\vspace{1truecm}

\bigskip

\section{The quark  potential from lattice QCD.}
The strong interactions of quarks, the fundamental constituents
of hadrons, are governed by the  Quantum
Chro\-mo\-dy\-na\-mics (QCD), the non-Abelian
gauge theory with the gluons as gauge bosons.
As a consequence of the non-Abelian character of QCD,
the quark interactions run from  one- to many gluon 
exchanges over gluon self-int\-er\-act\-ions, the latter being 
responsible for the 
so-called quark confinement, where highly energetic quarks remain trapped but 
behave as (asymptotically) free particles at high energies and momenta. The 
QCD equations are nonlinear and complicated due to the gluon 
self-int\-er\-act\-ion
processes and their solution requires employment of highly sophisticated 
techniques such as discretization of space time, so-called lattice QCD. 
Lattice QCD calculations have established themselves as a
reliable tool for non-per\-tur\-ba\-tive analysis of QCD.
The outcome (in the quenched approximation)
is a linear confinement potential with 
energy increase, be it quark--anti-quark $(Q\bar Q)$,
\begin{equation}
V_{Q\bar Q}(|\mathbf{r}|)=-\frac{A_{Q\bar Q}}{|\mathbf{r}|} +
\sigma_{Q\bar Q}|\mathbf{r}| +C_{Q\bar Q},
\label{Q_barQ}
\end{equation}
 or two-body potential between quarks (so called $\Delta $ type),
\begin{equation}
V_{3 Q}(\mathbf{r}_1,\mathbf{r}_2,\mathbf{r}_3)=
-A_{\Delta }\sum_{i<j}\frac{1}{|\mathbf{r}_i-\mathbf{r}_j|} +\sigma_{\Delta }
\sum_{i<j}|\mathbf{r}_i-\mathbf{r}_j|  +C_{\Delta },
\label{Q_Q}
\end{equation}
in obvious notations.
The Coulomb-like piece is associated with short range one-gluon exchange
in the perturbative regime,
whereas the long-range linear part relates to non-perturbative effects
and is attributed  to flux-tube 
$Q\bar Q$, or $QQ$ links. Its strength is then associated with the 
respective string tension \cite{Lattice_book}.
Detailed analysis of the values of the constants of $V_{Q\bar Q}$ and
 $V_{3Q}$ has been performed in Ref.~\cite{Takh}.
The three-quark $(3Q)$ Coulomb- plus linear potential, or, versions of it,
has found repeatedly application to  baryon spectroscopy \cite{Gonz,spto}.
In view of this, its generalization to an exactly solvable potential
is of interest.

We begin with  first drawing attention to the proximity of  
the two-body Coulomb- plus linear
potential (be it for $Q\bar Q$, $QQ$, or  $Q(QQ)$ systems), 
to  $(-\cot |\mathbf{r}| ) $.
Indeed, this is immediately seen from the corresponding Taylor expansion,
\begin{equation}
-\cot |\mathbf{r}| \approx -\frac{1}{|\mathbf{r}|} +
\frac{1}{3}\,|\mathbf{ r}| .
\label{Taylor_lin_HO}
\end{equation}
This expression shows that the absolute va\-lu\-es of the 
stre\-ngths of the linear to
Coulomb  potentials are in ratio 1:3,
a value that fits quite reasonably into  the range  of the 
$\sigma_{\Delta}:A_{\Delta}$
ratio of approximately $1/4- 1/2$ reported by the
lattice QCD analysis \cite{Takh}. 
In fact,  $(-\cot |\mathbf{r}|)$  is part of the  more general and 
exactly solvable 
trigonometric potential 
\begin{equation}
 v_{tRM}(|\mathbf{z}|)=-2 b \cot |\mathbf{ z}| +
a(a+1)\csc^2 \, |\mathbf{z}|\, ,\quad \mathbf{z}=\frac{\mathbf{r}}{d},
\label{v-RMt}
\end{equation}
known in supersymmetric quantum mechanics (SUSYQM) under the name of
the trigonometric Rosen-Morse potential \cite{DeDutt} and displayed  
in Fig.~1. The potential in eq.~(\ref{v-RMt})
is given in terms of dimensionless argument and parameters.
The figure shows that $v_{tRM}$ interpolates between the 
Coulomb-like potential (associated with the one-gluon exchange)
and the infinite well. The latter provides the adequate scenario
for the asymptotic freedom in so far as it describes trapped but free 
particles. The intermediate region of the by and large linear 
confinement potential can be attributed, as usual, 
to gluon flux tube links between the quarks. 
The length scale, $d$, will be considered as a free parameter 
to fit data.
The Taylor expansion of the $\csc^2$ term reads,
\begin{equation}
a(a+1) \csc^2 |\mathbf{z}| \approx 
\frac{a(a+1)}{\mathbf{z}^2} +
\frac{a(a+1)}{15}\mathbf{z}^2.
\label{Taylor_nlin_HO}
\end{equation}
The parameter  $a$ can be viewed as relative angular momentum
in the case when it takes integer non-negative values,
a para\-me\-tri\-za\-tion used in ref.~\cite{com06b}.
If so, the $\csc^2$ term acquires meaning of a non-standard 
centrifugal barrier. 

\begin{figure}
\resizebox{0.40\textwidth}{6.5cm}
{\includegraphics{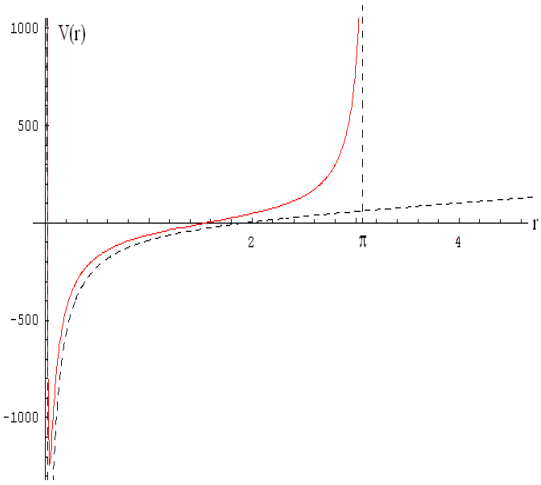}}
\caption{ The trigonometric Rosen-Morse potential (solid line) and its
proximity to the Coulomb-- plus linear potential (dashed line)  for the
toy values $a=1, b=50$ of the parameters.}
\label{fig_RMT}
\end{figure}

We were able to trace back $v_{tRM}$ to ref.~\cite{Levai}
where it apparently has been found  for the first time
as a pure mathematical construct while
completing the SUSYQM compilation of shape invariant
potentials that are exactly solvable by means of the hypergeometric 
differential equation. The $v_{tRM}$ potential is therefore one of the
new potentials predicted by SUSYQM \cite{DeDutt}.
We here suggest that the $QQ$ interactions are ruled by the
following trigonometric quark confinement (TQC) potential
\begin{eqnarray}
V_{3Q}^{\mathrm{TQC}}(\mathbf{z}_1,\mathbf{z_2}, \mathbf{z_3})&=&
-2b\sum_{i<j}
 \cot | \mathbf{z}_i-\mathbf{z}_j|\nonumber\\ 
+a(a+1)\sum_{i<j}\csc^2 |\mathbf{z}_i-\mathbf{z}_j|,
&\quad& \mathbf{z}_i=\frac{\mathbf{r}_i}{d}.
\label{TQC_QQ}
\end{eqnarray}
The great advantage of the TQC  potential
 over the Cou\-lomb- plus linear  potential
(compared in  Fig.~(\ref{fig_RMT})) is that while the latter is
neither especially symmetric, nor exactly solvable, the former is both,
it is exactly solvable,and  has in addition
the dynamical $O(4)$ symmetry \cite{com06b} for $a=l=0,1,2,... $,
and  $SO(2,1)$ for more general $a$ values \cite{Asim}. 

\section{The exact single particle basis of $v_{tRM}$.}
In this section we briefly review for the sake of completeness of the 
presentation  the exact solutions of the three dimensional
single particle
Schr\"o\-ding\-er equation with $v_{tRM}$,
as it appears in the quark-diquark approximation to nucleon structure
~\cite{com06b,com06a}.
From here onward we shall identify the $a$ parameter with the 
relative quark-diquark  angular momentum.
The above equation is solved in the usual 
way in se\-pa\-ra\-ting variables using the ansatz
\begin{equation}
\Psi_{(K l m)}(\mathbf{ z}) = 
Y_{(l m)}(\theta,\varphi){{\mathcal R}_{(K l)}(|\mathbf{z}|)\over
|\mathbf{ z}|}\ .
\label{wafu}
\end{equation}
The exact solutions of the radial part 
have been constructed in ref.~\cite{com06b}  on the basis of the 
one-di\-men\-sio\-nal solutions found in \cite{com06a} as:
\begin{eqnarray}
{\mathcal R}_{(K l)}(|\mathbf{z}|) =  N_{( K l) } 
\sin^{K+1} |\mathbf{z}| e^{-\frac{b|\mathbf{z}|}{K+1}  }
R_{K-l }^{(\frac{2b}{K+1 },-( K+1) )}&&\left(\cot |\mathbf{z}| \right),
\nonumber\\
K=0,1,2,...,\quad  l=0,1,...,K .&&
\label{Rom_pol}
\end{eqnarray}
Here,  $N_{(K l)}$ is a normalization constant and
$\mathbf{z}$ is the relative quark-diquark distance. 
The
$R_{n}^{(\alpha, \beta )}(\cot |\mathbf{z}|)$ 
functions are the non-classical  
Romanovski polynomials \cite{routh,rom}  which are defined by the
following Rodrigues formula,
\begin{eqnarray}
R_{n}^{(\alpha, \beta )}(x)&=&e^{\alpha  \cot^{-1} x} (1+x^2)^{-\beta +1}
\nonumber\\
&&\times \frac{d^n}{dx^n}e^{-\alpha  \cot^{-1} x} (1+x^2)^{\beta -1 +n},
\label{Rodrigues}
\end{eqnarray}
where $x=\cot |\mathbf{z}|$ (see ref.~\cite{raposo} for a recent review).
The  energy spectrum of the TQC potential is calculated as
\begin{eqnarray}
\epsilon_{K}=-b^2\frac{1}{(K+1)^2} +(K+1)^2,
&\,\,\,& K=n+l,
\label{spectr_RM}
\end{eqnarray}
where $n=0,1,2,...$. Therefore, particular quark levels 
bound within different TQC potentials (distinct by the values of $l$) 
happen to carry same energies and align to series of states (multiplets)
characterized by the superior quantum number $K$ (see Fig.~2).
One easily recognizes that the energy in eq.~(\ref{spectr_RM}) is
defined by the Balmer term and its inverse of opposite sign, thus
revealing $O(4)$ as dynamical symmetry of the problem.
Correspondingly, the  $K$-series  belong to the irreducible  
$O(4)$ representations of the type  $\left(\frac{K}{2},\frac{K}{2} \right)$ 
and $K$ has the 
meaning of  four-dimensional angular momentum.
Upon coupling the quark spin to them, one finds
the reducible $O(4)$ multiplet   
$\left(\frac{K}{2},\frac{K}{2} \right)\otimes 
\left(\frac{1}{2},0 \right)\oplus \left(0,\frac{1}{2} \right)$,
which consists of the $K$ parity mates
$\frac{1}{2}^\pm $, ...,$\left( K-\frac{1}{2}\right)^\pm$, 
and  the one unpaired state
of maximal spin $J_{\mathrm{max}}=K+\frac{1}{2}$
of either positive, or, negative parity.
\begin{figure}
\resizebox{0.40\textwidth}{6.5cm}
{\includegraphics{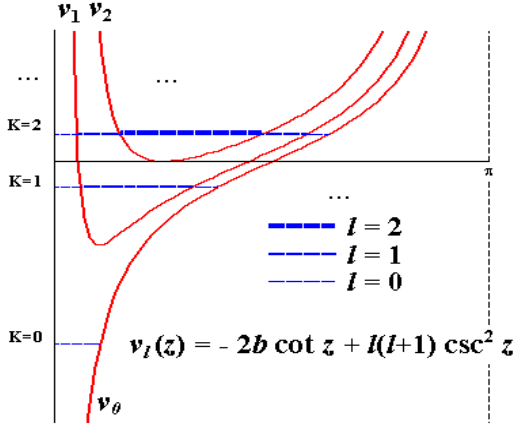}}
\caption{Schematic presentation of 
the lowest $K$ levels in eq.~(\ref{Rom_pol}).
They unite states of same energies from different 
TQC potentials, here denoted by
$v_l$,  whose  angular momenta vary
according to the rule $l=0,1,2,...,K$. }
\label{RoMoK}
\end{figure}
\subsection{The nucleon spectrum.}
Comparison of the TQC spectrum in Fig.~2 to the nucleon
excitations reported by \cite{PART}
reveals an amazing coincidence.
All the observed nucleon resonances with masses below $2.5$ GeV
do indeed fall into  $K=1,3,5$ multiplets
from which only the $F_{17}$ and $H_{1,11}$ states are still ``missing'',
an observation due to refs.~\cite{MK-97}.
The $K=2,4$ levels have been  attributed to entirely  ``missing'' resonances.
The unnatural parity of the $K=3,5$ states would require to account
of the internal structure of the diquark. 
This phenomenon can be interpreted as 
dominance of a quark-diquark configuration in nucleon
structure (see last reference in \cite{MK-97} for more details)
meaning that (i)  eq.~(10) can be directly employed to fit data on the
nucleon spectra, (ii) matrix elements of transition operators can
be evaluated in the basis of eq.~(8). 
The nucleon spectrum is fitted by the following potential parameters
\cite{com06b} 
\begin{eqnarray}
& b=5.85 \ ,\quad d=2.31 
\ \mathrm{fm}\, \quad \mu= 1.06 \,\, \mathrm{fm}^{-1},
\label{parameters}
\end{eqnarray}
where $\mu$ stands for the reduced mass of the quark --diquark system.
The wave function of the nucleon ground state, $K=0,l=0$, is obtained as 
\begin{eqnarray}
{\mathcal R}_{(00)}(|\mathbf{z}|)=N_{0}
e^{-b  |{\mathbf z}|\  }\sin |{\mathbf{z}|}, &\, &
N^2_{0}=\frac{4b (b^2+1)}{ 1-e^{-2 \pi b }}.
\label{wafu_gst}
\end{eqnarray}
In order to illustrate the predictive power of the TQC potential
and the efficiency of its exact single particle basis  
we shall exploit in the next section the potential parameters
in eq.~(\ref{parameters}) and the wave function in eq.~(\ref{wafu_gst})
in the calculation of the proton electric form-factor in 
the quark--diquark picture of baryon structure.
\section{The proton charge form factor.}
The electric form factor is defined in the standard way 
\cite{Roberts_rev} as
the  matrix element of the charge component,
$ J_0(\mathbf{r})$, of the proton electric current 
between the states of the incoming, $\mathbf{p}_i$, and 
outgoing, $\mathbf{p}_f$, electrons in the dispersion process,  
\begin{equation}
G_{\mathrm{E}}^{\mathrm{p}}(|\mathbf {q}|)=
<\mathbf{p}_f| J_0(\mathbf{r})|\mathbf{p}_i>,
\quad \mathbf{q}=\mathbf{p}_i-\mathbf{p}_f.
\label{el_ff}
\end{equation}
The explicit evaluation of eq.~(\ref{el_ff}), with
$J_0(\mathbf{r})=e_p|\psi_{\mathrm{gst}}(\mathbf{r})|^2$,
$e_p=1$, and plane waves for the electron states
 amounts to the calculation of the following integral,
\begin{eqnarray}
G_{\mathrm{E}}^{\mathrm{p}}(|\mathbf {q}|)
&=&\int_0^{\pi d} \mathrm{d}|\mathbf{r}| 
{({{\mathcal R}_{(00)}(|\mathbf{r}|)})^2 \sin{|\mathbf{q}||\mathbf{r}|}\over 
|\mathbf{q}||\mathbf{r}|}\ .
\label{el_ff_gst}
\end{eqnarray}
In the  $\int_0^{\pi d}\longrightarrow \int_0^\infty $ limit,
the integral calculates exactly and is given in terms of
$\widetilde \mathbf{q}=\mathbf{q}d $ as
\begin{eqnarray}
G_{\mathrm{E}}^{\mathrm{p}}(|\widetilde \mathbf {q}|)
&=& \frac{b (b^2+1)}{
|\widetilde{\mathbf{ q}}|} 
\nonumber\\
\times 
\tan^{-1}&& 
\frac{16 |\widetilde\mathbf{q}| b}{  \widetilde{\mathbf{q}}^4 
+ 4 (2 b^2 - 1) \widetilde{\mathbf{q}}^2 +16 b^2(b^2+1)}.
\label{FF}
\end{eqnarray}
The exact electric form factor of the proton obtained 
from eq.~ (\ref{FF}) is displayed
in Fig.~3 and follows pretty well the experimental data. 
\begin{figure}
\resizebox{0.40\textwidth}{6.5cm}
{\includegraphics{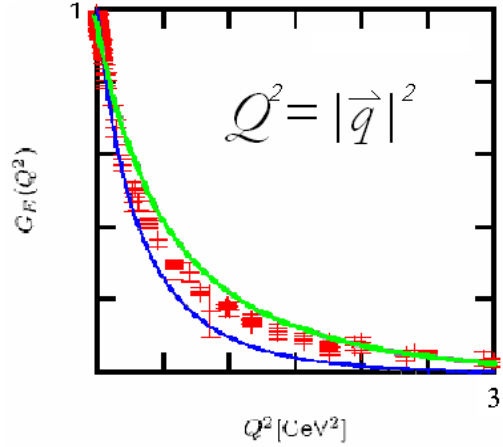}} 
\caption{Comparison of the  proton electric form factor
as obtained from the analytic expression in
eq.~ (\ref{FF}) (upper curve), and the numerical solution of 
the Bethe-Salpeter equation
(lower curve ) reported in ref.~\cite{haupt}.
In elastic electron-proton scattering, 
$Q^2=-q^2=-q_0^2+\mathbf{q}^2$, and $q_0=0$. 
The experimental data set is same as in \cite{haupt}.}
\end{figure}
\section{Effective gluon propagator.}
The proximity of the TQC potential to the QCD
quark-gluon dynamics is suggestive of the idea to
exploit the Born approximation
and introduce an effective gluon propagator 
as a Fourier  transform of that very potential.
In so doing one encounters the integral
\begin{eqnarray}
\Pi (|\mathbf{q}| )&=&
-\frac{1}{4\pi }\frac{2\mu }{\hbar^2} \int \mathrm{d}^3\mathbf{r} 
e^{i\mathbf {q}\cdot \mathbf{r}} V(\mathbf{r} ),
\end{eqnarray}
which, unfortunately diverges. As a remedy, we 
consider instead the  Fourier transforms of
potential matrix elements and find 
integrals that can be taken in closed form.
Specifically, the Fourier transform of the ground state matrix element
is calculated as
\begin{eqnarray}
\Pi^{(\mathrm {gst})}(|\mathbf{q}|)
&=&\int_0^{\pi d} \mathrm{d}|\mathbf{r}|\frac{\sin
|\mathbf{q}||\mathbf{ r}|}{
|\mathbf{q}||\mathbf{r}|} 
v_{tRM}(|\mathbf{r}|)\psi^2_{\mathrm{gst}}(\mathbf{r})
\mathbf{r}^2 \,\nonumber\\
=N^2_{(0,0)}\int _{0}^{\pi }{\mathrm{d}y} &&{\frac {{e^{-2\,b y}}
 \sin^2  y   \sin  q'y 
 \cot  y  }{q'y}}\ ,
\end{eqnarray}
 which expresses in terms of  
$Ei(u)=\int_1^\infty \frac{e^{-ut}}{t} \mathrm{d}t$, and
 $Ci (u)=\gamma +\ln u +\int_0^u \mathrm{d}t\frac{\cos t -1 }{t}$
functions, where
\begin{eqnarray}
y\equiv |{\mathbf z}|=\frac{|\mathbf{r}|}{d}, &\quad & 
q'\equiv |\widetilde{{\mathbf q}}|=\sqrt{Q^2}\, d\ .
\end{eqnarray}
In Figs.~4,5 we display $\Pi^{(\mathrm {gst})}(|\mathbf{q}|)$
for various values of the $b$ parameter.
The convenience of our consideration is supported by the
great similarity in the asymptotic behavior of the Coulomb
propagator and $\Pi^{(\mathrm {gst})}(|\mathbf{q}|)$.
\begin{figure}
\resizebox{0.40\textwidth}{6.5cm}
{\includegraphics{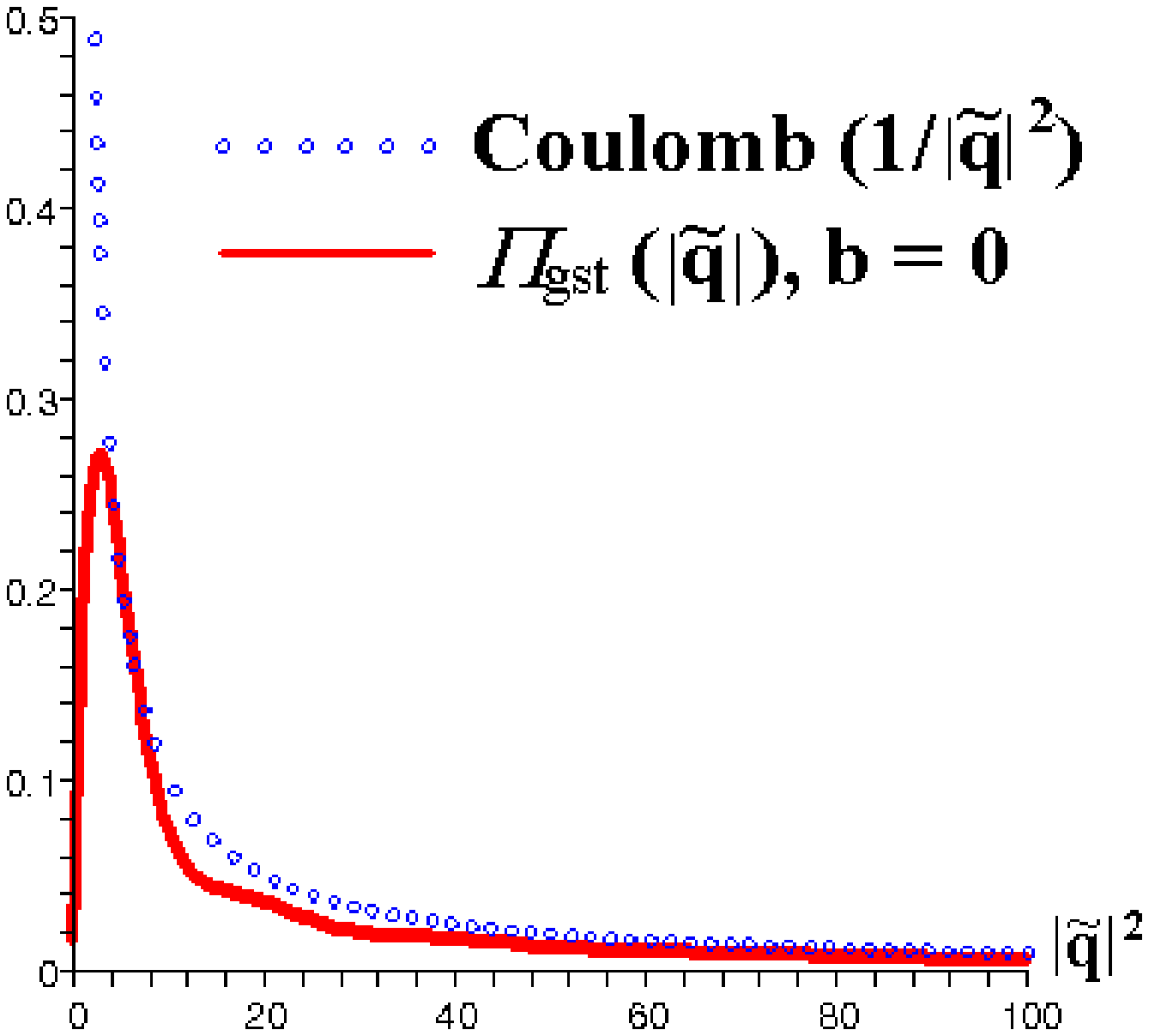}} 
\caption{Comparison of the Coulomb propagator (dotted line) with
the TQC-potential propagator for $b=0$ and normalized
ground state wave function (solid line).}
\end{figure}
\begin{figure}
\resizebox{0.40\textwidth}{6.5cm}
{\includegraphics{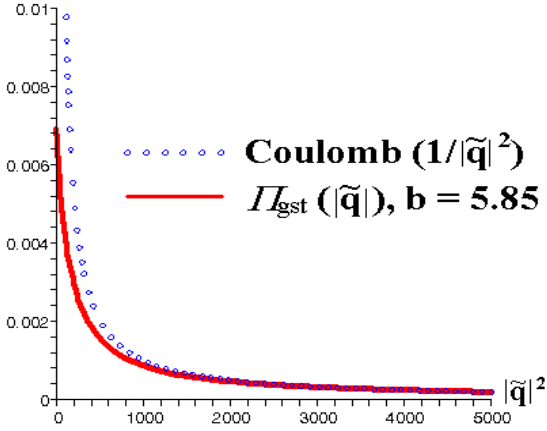}} 
\caption{Comparison of the Coulomb propagator (dotted line)
with the TQC-potential propagator for $b=5.85$ and unnormalized
ground state wave function 
(solid line).}
\end{figure}
\section{Concluding remarks.}
In this work we pointed out  that the trigonometric Rosen-Morse
potential captures quite realistically  
the traits of the QCD quark-gluon dynamics.
This is reflected by the facts that (i) the ratio
of $(-1/3)$ of the strengths of the linear to
Coulomb term following from the Taylor expansion of
the TQC potential is in line with the
value of $(-1/2)- (-1/4)$ calculated by lattice QCD, 
(ii) the prediction of the proton electric form factor in closed form
follows quite satisfactory data,
(iii) the nucleon spectrum is well reproduced
(and  the $\Delta $ one as well, skipped out here), 
(iv) the effective gluon propagator obtained as the Fourier
transform of the ground state matrix element  
of the TQC potential has reasonable shape and asymptotic behavior.
The exact single particle basis
of the TQC potential seems quite efficient, indeed, and provides
a good starting point for more detailed spectroscopic studies.
The TQC potential can be enriched, for example,  by the inclusion of 
$\mathbf{\sigma }_i\cdot\mathbf{ \sigma}_j$ interactions to account for the 
spin dynamics, and by screening effects
along the line of, say, ref.~\cite{Gonz}. 
The effective gluon propagator can be employed in studies of non-perturbative
QCD phenomena in the spirit of ref.~\cite{Roberts}.
In conclusion, the single particle basis of the TQC potential
considered here seems to provide an efficient tool for quark model
calculations of spectroscopic characteristics of baryons.


\end{document}